\documentclass[prd,nofootinbib,floats,superscriptaddress,eqsecnum,tightenlines,11pt]{revtex4}

\usepackage{hyperref}
\usepackage{graphicx}
\usepackage{amsmath,amssymb,amsfonts,amsthm,latexsym,stmaryrd}
\usepackage{marginnote}
\usepackage{color}
\usepackage{soul}

\newcommand{\beq}{\begin{equation}}
\newcommand{\eeq}{\end{equation}}

\newcommand{\U}{u}

\newcommand{\tb}{b}
\newcommand{\ptwo}{\sigma}

\begin{document}

\title{Neutron stars in degenerate higher-order scalar-tensor theories: Axial perturbations}
\author{Hamza Boumaza}
\affiliation{Laboratoire de Physique des Particules et Physique Statistique (LPPPS),\\
 Ecole Normale Supérieure-Kouba, B.P. 92, Vieux Kouba, 16050 Algiers, Algeria}
\author{David  Langlois}
\affiliation{Université Paris Cité, CNRS, AstroParticule et Cosmologie, F-75013 Paris, France}
\date{\today}

\begin{abstract}
We study the axial (or odd-parity) perturbations of neutron stars in a one-parameter subclass of Degenerate Higher-Order Scalar-tensor (DHOST) theories. After recalling  the equilibrium neutron star configurations obtained in a previous work  by solving the generalised Tolman-Oppenheimer-Volkoff equations in DHOST theories, we derive the action at  quadratic order in linear perturbations of axial type. We then compute the quasi-normal modes (QNMs) for several values of the modified gravity parameter and various equations of state, observing deviations of both frequencies and damping times  with respect to general relativity. We also analyze the impact of our modified gravity parameter on the universal relations relating the (rescaled) frequencies and damping times to the compactness of the neutron star.
\end{abstract}

\maketitle

\section{Introduction}
Compact objects, such as black holes and neutron stars, provide strong-gravity environments where potential deviations from general relativity could be expected to be the most significant,  in contrast with weak-gravity environments such as the Solar system. 
 In the present article, we focus our attention on  neutron stars in  scalar-tensor theories of gravity, more precisely in the context of degenerate higher-order scalar-tensor  (DHOST) theories \cite{Langlois:2015cwa,Langlois:2015skt,BenAchour:2016fzp} (see also \cite{Langlois:2018dxi,Kobayashi:2019hrl} for reviews).

 In a previous work \cite{Boumaza:2022abj},  
 we have  studied neutron star configurations in a single-parameter  shift-symmetric subfamily of   DHOST theories. Assuming static, and thus spherically symmetric, neutron stars, we computed  their radial profile for  several realistic equations of state describing  the properties of nuclear matter.  We found that,  in these modified gravity theories, neutron stars could have masses and radii significantly larger than their GR counterparts.
 We also extended our results to slowly-rotating solutions, establishing relations between the dimensionless moment of inertia and the compactness of the star \cite{Breu:2016ufb}, relations  that are quasi universal  in the sense that they depend only weakly on the choice of  the equation of state. Neutron stars have also been studied in other particular subfamilies of DHOST theories (see e.g. \cite{Cisterna:2016vdx,Maselli:2016gxk,Minamitsuji:2016hkk,Babichev:2016jom,Sakstein:2016oel,Kobayashi:2018xvr,Chagoya:2018lmv,Boumaza:2021hzr,Kase:2021mix,Boumaza:2021lnp,minamitsuji2022stability}). 

 Beyond static solutions, it is also important to study the linear perturbations of neutron stars  in modified gravity, in order to investigate their stability, as well as identify their oscillatory modes, known as quasi-normal modes (QNMs). In particular, QNMs are expected to be excited in a newly formed  compact object resulting from the merger of two  compacts objects in a binary. Such event possesses three phases: inspiral, merger and ringdown, during which gravitational waves are emitted. The  signal in the ringdown phase  can  be decomposed into, mostly, quasi-normal modes, which depend on the physical characteristics of the compact object, as well as on the model of gravity. Analyzing the QNM spectrum of neutron stars  thus not only gives  information about the nuclear equation of state but also constrains  theories of gravity. 

In practice, the dynamics of the linear perturbations can be obtained by expanding the total action, 
i.e. the sum of  the gravitational action and of the matter action, up to second order in perturbations, as  was performed, for instance, in \cite{Kase:2021mix} for neutron stars in Horndeski theories. 
For a static and spherically symmetric background, the  perturbations can be decomposed  into odd-parity, or axial, modes and even-parity, or polar,  modes. 
In the present work, we concentrate on the axial modes   of the neutron stars studied in \cite{Boumaza:2022abj}. Axial perturbations, as they are odd-parity, do not couple to the fluid and scalar field perturbations. They obey an equation of motion  that generalises the well-known Regge-Wheeler equation of GR. In fact, we show that this equation is the same as that obtained in GR with another metric, dubbed effective metric and which can be deduced from the background metric by a disformal metric. A similar correspondence between  DHOST theories and GR  has been studied recently for black holes~\cite{Langlois:2022ulw}.

The outline of our paper is the following. In Section \ref{SEC1}, we introduce the general framework of DHOST theories and specify the subclass under consideration. In Section \ref{SEC2}, we construct static and spherically symmetric neutron star backgrounds by solving the modified TOV equations. In Section \ref{SEC3}, we perform a second-order perturbative expansion of the action to derive the equation of motion for axial modes. Section \ref{SEC4} is devoted to the numerical computation of the fundamental $l=2$ quasinormal mode and of the impact of DHOST corrections on the universal relations. We conclude in Section \ref{conclusion} with a summary of our findings and discuss possible extensions of this work.
Additional details are provided in two appendices.

\section{Gravity models}\label{SEC1}

In this section, we briefly recall the modified gravity models that we consider. DHOST theories~\cite{Langlois:2015cwa,Langlois:2015skt,BenAchour:2016fzp} represent the most general scalar-tensor theories propagating a single scalar degree of freedom, and include, as particular subfamilies,  the traditional scalar-tensor theories,  Horndeski theories~\cite{horndeski1974second} and so-called Beyond Horndeski theories~\cite{Gleyzes:2014dya,Gleyzes:2014qga}. \\
Up to quadratic order (in second derivatives of the scalar field), their action is of the form 
\begin{equation}
S_{\rm grav}[g_{\mu\nu},\varphi]=\int d^4x\sqrt{-g}\left(K+Q\nabla^{\mu}\nabla_{\mu}\varphi+F\,R+\sum_{i=1}^{5}A_{i}L_{i}\right)\label{Stot}
\end{equation}
where $R$ is the Ricci scalar associated with the metric $g_{\mu\nu}$,
the $L_i$ denote
the five elementary Lagrangians  that are quadratic in second order derivatives of the scalar field  (contracted with the metric or the scalar field gradient), explicitly given by 
\begin{eqnarray*}
L_{1}=\nabla^{\nu}\nabla^{\mu}\varphi\nabla_{\nu}\nabla_{\mu}\varphi & L_{2}=\left(\nabla^{\mu}\nabla_{\mu}\varphi\right)^{2} & L_{3}=\nabla^{\alpha}\nabla_{\alpha}\varphi\nabla_{\nu}\nabla_{\mu}\varphi\nabla^{\nu}\varphi\nabla^{\mu}\varphi\\
 & L_{4}=\nabla^{\nu}\varphi\nabla_{\nu}\nabla_{\mu}\varphi\nabla^{\mu}\nabla^{\alpha}\varphi\nabla_{\alpha}\varphi\,, \qquad
 & L_{5}=\left(\nabla^{\nu}\varphi\nabla_{\nu}\nabla_{\mu}\varphi\nabla^{\mu}\varphi\right)^{2}\,,
\end{eqnarray*}
and the functions $K$, $Q$, $F$ and 
$A_{i}$  depend in general  on the scalar field $\varphi$ and on its asssociated kinetic term $X\equiv\nabla^{\mu}\varphi\nabla_{\mu}\varphi$. In the following, we will consider shift-symmetric theories where these functions depend on $X$ only. Moreover, whereas the functions $K$, $Q$ and $F$ are arbitrary, three of the functions $A_i$ must be  related to the other two via the so-called degeneracy conditions, given explicitly in \cite{Langlois:2015cwa}, so that  the theory is degenerate and thus contains a single scalar degree of freedom. 

As in our previous work \cite{Boumaza:2022abj},  we restrict the present study to a subfamily of DHOST theories, characterised by
the choice of functions
\beq
\label{A1_A2}
K=Q=A_{1}=A_{2}=A_5=0\,,\qquad 
A_{3}=-A_{4}=-4\frac{F_X}{X}\,,
\eeq 
which satisfy the degeneracy conditions mentioned above.
The corresponding action,
\begin{equation}
S_{\rm grav}=\int d^{4}x\, \sqrt{-g}\left(F\,R-4\frac{F_X}{X}(L_3-L_4)\right)\,,
\label{S}
\end{equation}
thus depends on the single function $F(X)$.
The above action belongs to the 'Beyond Horndeski' subfamily of DHOST theories, introduced in  \cite{Gleyzes:2014dya,Gleyzes:2014qga}.
If $F$ is a constant, then one recovers general relativity.
Note that the condition $A_1=A_2=0$ implies  that the propagation speed of  gravitational waves is strictly  equal to that of light~\cite{Langlois:2017dyl}. 

To take into account the matter in the interior of the neutron star, 
the gravitational action  must be supplemented by 
 an  action representing the dynamics of the matter, which is assumed to be minimally coupled to the metric $g_{\mu\nu}$. In general, matter in neutron stars is described by a perfect fluid, with pressure $P$ and energy density $\rho$. Although the corresponding energy-momentum tensor,
 \begin{eqnarray}
 \label{T}
T_{\alpha\beta}=(P+\rho)u_\alpha u_\beta + P g_{\alpha\beta}\,,
\end{eqnarray}
 is very simple, an action principle for a perfect fluid is more tricky. Several actions have been proposed in the literature \cite{Taub:1954zz,schutz1970perfect,ray1972lagrangian,schutz1977variational,carter1989relativistic,brown1993action}
 and we will adopt here the action~\cite{schutz1970perfect}
\begin{eqnarray}
S_{\rm m} =\int d^4 x \sqrt{-g}\,P(\mu) , \label{action}
\end{eqnarray}
 where  $P(\mu)$  denotes the pressure  as a function of the chemical potential $\mu$,  which defines the matter equation of state\footnote{For a neutron star equation of state, one can ignore the thermal effects and consider an equation of state of the form $P(\mu)$ instead of the more general equation of state $P(\mu,s)$, where  $s$ is the entropy density.}. The scalar $\mu$ is given by the norm of the covector $\mu_\alpha$ (i.e. $\mu=\sqrt{-g^{\alpha\beta}\mu_\alpha\mu_\beta}$), assumed to be proportional to the fluid four-velocity $u_\alpha$ and defined in the above variational principle as
 \begin{equation}\label{decompositionofu}
     \mu_\alpha=\mu \, u_\alpha=\partial_\alpha L + A\partial_\alpha B\,,
 \end{equation}
 where $L$, $A$ and $B$ are three scalar fields. By varying the action \eqref{action} with respect to these three scalar fields,  one  recovers the usual equations of motion for a perfect fluid, as explained in the Appendix.

\section{Background equations}\label{SEC2}
We now describe our background configuration. 
For a non-rotating star, the metric is static and  spherically symmetric, i.e. of the form
\begin{equation}
ds^{2}=-f(r)dt^{2}+h(r)dr^2+r^2\left(d\theta^2+\sin^2\!\theta\,  d{\phi}^2\right)\,,
\label{ds}
\end{equation}
and the scalar field  can be written as
\beq
\label{phi}
\varphi(t,r)=q\;t+\psi(r)\,,
\eeq
where $q$ is a real constant. This is the most general form for the scalar field compatible  with a static and spherically symmetric geometry in a shift symmetry of the theory: indeed, only the gradient of the scalar field, which is time independent, appears in the equations  of motion.

Substituting the above metric (\ref{ds}) and scalar field (\ref{phi}) into the total action $S=S_{\rm grav}+S_{\rm m}$, we get, after integration by parts, the expression
\begin{equation}
S_{\rm bgd}=4\pi\int dr\,dt \left[\, r^2\sqrt{f h}\left(\frac{2 F h'}{h^2 r}+\frac{2 F (h-1)}{h r^2}-\frac{4 F_X X' \left(f X+q^2\right)}{f h r X}+ P \right)+ \lambda_0\, (\mu\sqrt{f})'\,\right]\,,
\label{S2}
\end{equation}
where a prime denotes a derivative with respect to the radial coordinate $r$ and   the kinetic term $X$ must be seen as a function of  the metric coefficients   in \eqref{ds} and of the scalar field components\eqref{phi}, namely
\begin{equation}
\label{X}
    X=\psi'^2/h-q^2/f\,.
\end{equation}
We have also added in the action an extra term, proportional to the Lagrange multiplier $\lambda_0$, that  enforces   the constraint
\begin{equation}
    \frac{\mu'}{\mu}= -\frac{f'}{2 f}\,,
\end{equation}
which follows from the fluid equation of motion in the background (see Appendix~\ref{Appendix_fluid}). In this way, instead of considering  the action as a functional of the three scalar fields $L$, $A$ and $B$ as discussed in the previous section, the action \ref{S2} is simply  a functional of $\mu$, of the metric coefficients $f$ and $h$ and of the scalar field $\psi$. 

 By varying the action (\ref{S2}) with respect to $f$ and $h$, we obtain the time and radial equations of motion which read, respectively,
\begin{eqnarray}
&&h' \left(\frac{F_{\text{X}} \left(-4 f^2 \left(\psi '\right)^4+8 f h q^2 \left(\psi '\right)^2+4 h^2 q^4\right)}{f h^2 q^2-f^2 h \left(\psi '\right)^2}+2 F\right)-\frac{8 F_{\text{X}} \left(f \left(\psi '\right)^3-2 h q^2 \psi '\right)}{f \left(\psi '\right)^2-h q^2} \psi '' \nonumber\\
&&+\frac{4 h q^2 F_{\text{X}} \left(f (h+1) \left(\psi '\right)^2-(h-1) h q^2\right)}{f r \left(f \left(\psi '\right)^2-h q^2\right)}+\frac{2 F (h-1) h}{r}-h^2 r \rho  =0,\label{eh}\\
&&f' \left(\frac{F_{\text{X}} \left(-4 f^2 \left(\psi '\right)^4+8 f h q^2 \left(\psi '\right)^2+4 h^2 q^4\right)}{f^2 h^2 q^2-f^3 h \left(\psi '\right)^2}+\frac{2 F}{f}\right)+\frac{8 q^2 F_{\text{X}} \psi ' \psi ''}{h q^2-f \left(\psi '\right)^2}\nonumber\\
&&+\frac{F_{\text{X}} \left(4 (h-1) h q^2 \left(\psi '\right)^2-4 f (h+1) \left(\psi '\right)^4\right)}{h r \left(h q^2-f \left(\psi '\right)^2\right)}+\frac{F (2-2 h)}{r}-h r P=0\,,
\label{ef}
\end{eqnarray}
where, in the first equation, we have used 
\begin{eqnarray}\label{lamdaprime}
\lambda_0' =    r^2  \sqrt{h} \, \frac{dP}{d\mu}\,,
\end{eqnarray}
which follows from the variation of  \ref{S2}  with respect to $\mu$, as well as the expression of the energy density $\rho$ in terms of $P(\mu)$,
\begin{equation}
\label{rho}
    \rho=\mu \frac{dP}{d\mu}-P\,.
\end{equation}

Finally, by varying  the action \eqref{S2} with respect to $\psi$, we obtain the scalar field equation of motion. In the shift-symmetric case, this is related to the conservation of a four-dimensional current, $\nabla_\mu J^\mu=0$, which reduces, due to the symmetries of the configuration, to 
\begin{eqnarray}\label{ephi}
\frac{d}{dr}\left(\sqrt{fh}\, r^2 J^r\right)=0\,,
\end{eqnarray}
with
\begin{eqnarray}
\label{Jr}
J^r=\frac{2 \sqrt{f X+q^2}}{f^{3} h^{3}r^2 X}\left\{-2 F_X \left[h r f' \left(f X-q^2\right)+f \left(f h (h+1) X-q^2 r h'+2 h q^2\right)\right]\right\},
\end{eqnarray}
which implies (to avoid a singularity at the center)
\beq\label{Jr0}
J^r=0\,.
\eeq
We must complete our set of equations with the equation of state $P=P(\mu)$, or equivalently $P=P(\rho)$. It can also be convenient to use the matter conservation equation $\nabla_\mu T^\mu_\nu=0$, which yields
\begin{eqnarray}
P'=-\frac{f'(P+\rho)}{2f}\,.
\label{e4}
\end{eqnarray}
This equation can also be derived from  \eqref{eh}, \eqref{ef} and the scalar field equation of motion \eqref{ephi}. 

In summary, for a given function $F$ characterising the theory of gravity \eqref{S} and some equation of state, the above equations form a radial differential system which can be integrated numerically, providing neutron star solutions determined by their central density or, equivalently, their mass. Such configurations of neutron stars were obtained in our previous work \cite{Boumaza:2022abj} for a one-parameter family of modified gravity models defined by the function 
\begin{equation}
    F=\frac{\kappa}{2}+\sigma X, 
\qquad 
\kappa\equiv (8\pi G)^{-1}\,,
\end{equation}
where the parameter $\sigma$ quantifies the deviation from GR, and for several parametrised realistic equations of state which will be recalled below in section \ref{SEC4}. In the following, we  will work in units where  $G=c=1$, so that $\kappa=(8\pi)^{-1}$.

\section{Axial perturbations}\label{SEC3}
We now turn to the study of axial perturbations (see  Appendix~\ref{Appendix_axial} for more details). It is convenient to decompose these perturbations into spherical harmonics $Y_{lm}(\theta,\phi)$, with coefficients that depend  on the coordinates $t$ and $r$.  The spherical symmetry of the background entails a degeneracy with respect to the value of $m$, so  we can restrict our analysis to the simplest value  $m=0$ without loss of generality and study the modes with distinct values of $l$ separately since they are decoupled at linear order.
Choosing the Regge-Wheeler gauge \cite{Regge:1957td}, we thus consider  the perturbed metric to be of the form
\begin{eqnarray}
\label{perturbed_metric}
ds^2&=& -f (r)dt^2+h(r)dr^2+r^2\left(d\theta^2 + \sin^2 \theta d\phi^2\right)\nonumber\\
& &+2 r^2\sin \theta \partial_\theta Y_{l0}(\theta)\left(H_0(r,t)dtd\phi +H_1(r,t)drd\phi \right),
\end{eqnarray}
with the two perturbations\footnote{We use the notation $H_0$ and $H_1$ instead of the more traditional notation $h_0$ and $h_1$ to avoid possible confusions with the metric coefficient $h$.} $H_0(r,t)$ and $H_1(r,t)$ depending implicitly on $l$.

 For axial perturbations, the gravitational scalar field is unperturbed, i.e. $\delta\varphi=0$, while  the perturbed four-velocity  of the fluid is of the form 
\begin{eqnarray}
\label{4-velocity}
u^\mu=\bar{u}^\mu+\delta u^\mu\,, \qquad \bar{u}^\mu=\frac{\delta^\mu_0}{f^{1/2}}\,,\qquad 
\delta u^\mu = \left\lbrace 0,0,0,v(r,t)\frac{\partial_\theta Y_{l0}(\theta)}{f^{\frac{1}{2}}\sin\theta  } \right\rbrace\,.
\end{eqnarray}

If we substitute the perturbed metric \eqref{perturbed_metric} and the perturbed four-velocity \eqref{4-velocity} into the energy-momentum tensor \eqref{T}, we find that the non-vanishing perturbed components of the tensor,  at linear order in the perturbations, are given by
\begin{eqnarray}
\label{T_tphi}
\delta T^{t\phi}&=&\frac{\partial_\theta Y_{l0}(\theta)}{\sin \theta\, f}\left[ P H_0+(\rho+P)v\right],\\
\label{T_ttheta}
\delta T^{r\phi}&=&-\frac{\partial_\theta Y_{l0}(\theta)}{\sin \theta\, h}P H_1.
\end{eqnarray}
Then, using the perturbed matter conservation equation, one can easily find that 
\begin{eqnarray}
\label{H0}
v=-H_0\,.
\end{eqnarray}
This shows that there is no independent degree of freedom associated with the fluid in the axial sector.

In order to derive the dynamics of the single degree of freedom associated with axial perturbations, it is convenient to expand the total  action up to  second order in the perturbations (see the appendices  for details). After integrating by parts, we get
\begin{eqnarray}
S_{\rm axial}=l(l+1)\int drdt \left[Q_2 H_0^2- Q_1 H_1^2 +Q_3 H_1 H_0+Q_0 ( H_0'-\dot{H}_1)^2  \right]\,,\label{Saxial}
\end{eqnarray}
where the coefficients of the quadratic terms are given by
\begin{eqnarray}
Q_0&=&\frac{F  r^4}{2 \sqrt{f\,h}},\qquad Q_1=\frac{\sqrt{f} F r^2 ( l (l+1)-2) }{2 \sqrt{h}},\qquad Q_2=\frac{h}{f}Q_1,\nonumber\\
Q_3&=&-\frac{2 \sqrt{f} \sqrt{h} q r^2 F_{\text{X}} \psi ' \left(f \left(\psi '\right)^2 \left(4 F+P r^2\right)-q^2 \left(4 F (h-1)+h r^2 (2 P+\rho )\right)\right)}{f F h \left(f \left(\psi'\right)^2-h q^2\right)+2 F_{\text{X}} \left(f^2 \left(\psi '\right)^4-2 f h q^2 \left(\psi '\right)^2-h^2 q^4\right)}\,,
\end{eqnarray}
evaluated on the background solution. Interestingly, one notes that $Q_3=0$ as a consequence  of the scalar field equation of motion (\ref{Jr}-\ref{Jr0}).

Given the structure  of the quadratic action, it is convenient to  introduce the
new variable 
\begin{eqnarray}
  \chi=\dot{H}_1 -H_0'\,.
\end{eqnarray}
Implementing it in the variational principle via a Lagrange multiplier $\lambda$, we consider  the equivalent action
\begin{eqnarray}
S'_{\rm axial}=l (l+1)\int drdt\,   \left[Q_2 H_0^2- Q_1 H_1^2 +Q_0 \chi ^2 +\lambda  \left(\chi + H_0'-\dot{H}_1 \right)\right]\,.
\label{Saxial}
\end{eqnarray}
 By varying (\ref{Saxial}) with respect to $H_1$, $H_0$ and $\chi$, we obtain
\begin{eqnarray}\label{eqaxial}
\lambda &= &- 2 Q_0 \chi ,\quad \lambda' = 2 Q_2 H_0,\quad \dot{\lambda} = 2 Q_1 H_1.
\end{eqnarray}

Note that there is no $l=0$ mode in  the axial sector.  When $l=1$ the coefficients $Q_1$ and $Q_2$ vanish and  the above equations imply 
\begin{eqnarray}
    \dot{\chi}=0,\quad (Q_0 \chi)'=0\,,
\end{eqnarray}
which can be integrated to give
\begin{eqnarray}
    \chi=\chi_0 \frac{\sqrt{fh}}{F r^4},
\end{eqnarray}
where $\chi_0$ is a constant of integration. This solution corresponds to a slowly-rotating star, as it can be seen by fixing the gauge $H_1=0$, so that the only metric perturbation is $H_0=-\int dr \chi$, which can be related to the star angular momentum (see e.g. our previous work \cite{Boumaza:2022abj}).
\\

For $l\geq 2$, by solving the equations \ref{eqaxial} with respect to $H_1$, $H_0$ and $\chi$ and then  inserting the results in the action (\ref{Saxial}), we obtain, after integration by parts, the action
\begin{eqnarray}\label{dS2}
S'_{\rm axial}=\int dr\, dt\, \frac{l (l+1) }{4Q_1 }\left[\dot{\lambda} ^2-\frac{f}{h}\lambda'^2 - \frac{Q_1}{ Q_0}\lambda ^2\right].
\end{eqnarray}
As expected, this action contains a single degree of freedom. 
Varying the action (\ref{dS2}) with respect to $\lambda$, we get the equation of motion
\begin{eqnarray}\label{pereq2}
   \ddot{\lambda} -Q_1\frac{d}{dr}\left[\frac{1}{Q_1 }\frac{f }{h}\lambda'\right] +\frac{Q_1}{ Q_0}\lambda =0\,.
\end{eqnarray}

Up to an irrelevant global factor, one can also rewrite the quadratic action \eqref{dS2} in the form
\begin{eqnarray}
    S''_{\rm axial}
    &=&
    \frac12\int dt dr_*\,  \frac{1}{Fr^2 }\left[\dot{\lambda} ^2-(\partial_*\lambda)^2 - \frac{Q_1}{ Q_0}\lambda ^2\right]\,,
    \label{dS2new}
\end{eqnarray}
where we have introduced the new radial coordinate $r_*$, defined by 
\begin{equation}
    dr_{*}=\sqrt{\frac{h}{f}}\, dr\,,
\end{equation}
and where the symbol $\partial_*$ denotes a partial derivative with respect to $r_*$. 
In terms of this new radial coordinate, one sees immediately that the propagation speed  of the axial perturbation is equal to that of light. Moreover, it is also clear that there is no ghost or gradient instability as soon as $F>0$.

Variation of the action \eqref{dS2new} yields the equation of motion 
\begin{equation}
\label{eom}
    \ddot\lambda-Fr^2\frac{\partial}{\partial r_*}\left[\frac{1}{Fr^2}\frac{\partial\lambda}{\partial r_*}\right]+\frac{f}{r^2}\left(l(l+1)-2\right)\lambda=0\,.
\end{equation}
Going into the frequency domain, via the Fourier transformation 
\begin{equation}
\lambda(t,r)=\int d\omega\,e^{i\,\omega\,t}\lambda(\omega,r)\,,
\end{equation}
and introducing the new function
\begin{eqnarray}
    U\equiv \frac{\lambda}{r\sqrt{F}}\,,
\end{eqnarray}
\eqref{eom} leads to a Schr\"odinger-like equation, of the form
\begin{equation}
-\frac{d^2 U}{dr_*^2} +V_l U=\omega^2   U\,,
\end{equation}
with the potential
\begin{equation}
\label{potential}
    V_l=\frac{f}{r^2}\left(l(l+1)-2\right)-{r\sqrt{F}}\,\partial_*\left(\frac{\partial_*(r\sqrt{F})}{{r^2F}}\right)\,.
\end{equation}

Interestingly, the above equation of motion can be obtained from the GR dynamics of  axial perturbations  in an {\it effective} metric given by   \begin{equation}
\label{effectivemetric}
\mathrm{d} \tilde{s}^2=
F\left[-fdt^{2}+h dr^2+r^2\left(d\theta^2+\sin^2\!\theta\,  d{\phi}^2\right)\right]\,.
\end{equation}
The situation here is analogous to axial perturbations of non-rotating black holes in DHOST theories, studied  in \cite{Langlois:2022ulw}. As shown there, the correspondence between the background metric and the effective metric can be understood, at least for quadratic DHOST theories, as a disformal transformation relating these two metrics (see \cite{BenAchour:2016cay,Langlois:2020xbc}  
 for disformal transformations in DHOST theories) so that the functions characterising the new theory, i.e. after disformal transformation,  satisfy $\tilde{F}=1$ and $\tilde{A}_1=0$. In our case, the initial theory already satisfies $A_1=0$, so one just needs here a purely conformal transformation, with the factor $F(X)$, which corresponds to the effective metric \eqref{effectivemetric}. 

\section{Numerical analysis}\label{SEC4}
In this section, we focus our attention on the theories \eqref{S} with 
\begin{equation}
    F=\frac{\kappa}{2}+\sigma X, 
\end{equation}
which we have considered in our previous paper \cite{Boumaza:2022abj}, and present the numerical analysis of the axial perturbations in this simple illustrative case. 

The numerical background solutions have already been computed in our previous work \cite{Boumaza:2022abj}, where we found that the amplitude of the beyond-GR effects on the neutron star depends only on the parameter combination
\begin{equation}
    p\equiv  q^2 \ptwo\,,
\end{equation}
which has the same dimension as $\kappa$.

We  considered five realistic equations of state, known as SLy, FPS BSk19, BSk20 and BSk21, discussed in  \cite{Haensel:2004nu,Potekhin:2013qqa}. They can  be parametrised in the form
\begin{eqnarray}
\log_{10}\left(\frac{P}{{\rm g\;cm}^{-3}}\right) &=& \frac{\tb_1+\tb_2 \xi +\tb_3 \xi ^3}{1+\tb_4 \xi}\U\left[\tb_5 (\xi -\tb_6)\right]+(\tb_7+\tb_8 \xi)\U\left[\tb_9 (\tb_{10}-\xi )\right]+(\tb_{11}+\tb_{12} \xi)\U\left[\tb_{13} (\tb_{14}-\xi )\right]\nonumber\\
& & +(\tb_{15}+\tb_{16} \xi )\U\left[\tb_{17} (\tb_{18}-\xi )\right]
 +\frac{\tb_{19}}{\tb^2_{20} (\tb_{21}-\xi )^2+1}+\frac{\tb_{22}}{\tb^2_{23} (\tb_{24}-\xi )^2+1},\label{EOS}
\end{eqnarray} 
with 
\begin{eqnarray}
\U[x]\equiv\frac{1}{e^x+1}\,,\qquad  \xi=\log_{10}({\rho}/g\;{\rm cm}^{-3})\,.
\end{eqnarray}
Each equation of state is characterised by the values of the coefficients $\tb_i$. 
 For the  SLy and FPS equations of state, these coefficients are given by
 \beq
 \tb_i=a_i^{\rm HP} \quad {\rm for }\quad 1\leq i \leq 18\,,\qquad  \tb_j = 0\quad {\rm for} \quad 19\leq j\leq 24\,,
 \eeq
 where the  $a_i^{\rm HP}$ denote the coefficients $a_i$ of \cite{Haensel:2004nu}.
For the BSk19, BSk20 and BSk21 equations of state,  the coefficients are
\beq
 \tb_i=a_i^{\rm PFCPG} \quad {\rm for }\quad 1\leq i \leq 9\,,\qquad \tb_{10}= a_6^{\rm PFCPG}\,, \qquad   \tb_j =a_{j-1}^{\rm PFCPG} \quad {\rm for} \quad 11\leq j\leq 24\,,
 \eeq
 where the  $a_i^{\rm PFCPG}$ correspond to  the coefficients $a_i$ of 
\cite{Potekhin:2013qqa}.

The background equations are numerically integrated from the center  to the surface of  the star,  for different value of the parameter $p$ and of the central energy density $\rho_c$.  At the center of the star, $r=0$, we impose the regularity  of the metric, energy density and scalar field. We then integrate from $r=0$ to $r=r_s$, where $r_s$ is the radius of the star where the pressure vanishes, i.e.  $P(r_s)=0$.

Outside the star, the metric is given by the usual Schwarzschild metric, i.e. 
\begin{eqnarray}
\label{Schwarzschild}
    f=\frac{1}{h}=1-\frac{2M}{r},
\end{eqnarray}
where the mass of the star $M$, determined from  the continuity of the metric at $r=r_s$, depends on the values  of $\rho_c$ and $p$.  As for the  scalar field  outside the star, it is determined  by solving Eq.~\eqref{Jr}  in the Schwarzschild geometry \eqref{Schwarzschild}, which implies $X=-q^2$ and therefore 
$F=F_{\rm ext}\equiv  \frac{\kappa}{2}-p$. The scalar field  profile  is then obtained by solving  \eqref{X} for $\psi$, which gives 
\begin{eqnarray}
\psi'=\frac{q\sqrt{2 M  r}}{r-2 M}\,,
\label{psi_ext}
\end{eqnarray}
which vanishes at spatial infinity.

Now, we turn our attention to the axial perturbation equation (\ref{pereq2}), which takes the following form in  the frequency domain:
\begin{eqnarray}\label{pereq22}
   Q_1\frac{d}{dr}\left[\frac{1}{Q_1 }\frac{f}{h}\lambda'\right] +(\omega^2-\frac{Q_1}{ Q_0})\lambda =0.
\end{eqnarray} 
We note that $\omega=\omega_R+i\,\omega_I$ is a complex quasinormal frequency where $\omega_R$ is interpreted as the  oscillation frequency, while $|\omega_I|$ is the inverse of the damping time $\tau$. Near the center of the star, the general solution of the above equation behaves as\footnote{By  doing a Taylor expansion near the center of the star, one finds that the leading order part of Eq.\eqref{pereq22} reads $$\lambda''-\frac{2 }{r}\lambda '-\frac{  (l^2+l-2)}{r^2}\lambda=0\,,$$ yielding \eqref{smalllambda}.}
\begin{eqnarray}\label{smalllambda}
    \lambda(\omega,r)\sim \lambda_-\, r^{1-l}+\lambda_+ \,r^{l+2}
    \qquad (r\to 0)
    \,,\end{eqnarray}
where $\lambda_+$ and $\lambda_-$ are constants of integration. 
Outside the star, the function $F$ takes the   constant value 
$F_{\rm ext}$. Moreover,  the potential $V_l$ defined in \eqref{potential} vanishes at spatial infinity, so that the solution at spatial infinity is, like in GR, of the form
\begin{eqnarray}\label{solinf}
  \lambda(\omega,r)\sim \lambda_{\rm in}e^{i \omega\, r_*}+\lambda_{\rm out}e^{-i \omega\, r_*}\qquad (r_*\to\infty)\,,
\end{eqnarray}
where $\lambda_{\rm in}$ and $\lambda_{\rm out}$ are also constants of integration. This solution is a linear combination of an outgoing and an ingoing wave. In order to have a purely outgoing wave and a regular solution at the center of the star, we have to set $\lambda_-=0$ and $\lambda_{\rm in}=0$. These boundary conditions are satisfied only for a discrete set of complex  values of $\omega$ corresponding to the  quasinormal modes.

In practice, due to the exponential growth of the outgoing wave at spatial infinity, while the ingoing wave is small, we are not sure to have a purely outgoing wave function in the numerical treatment. To overcome this problem, we follow the method presented in Ref.~\cite{andersson1995new} and define a complex coordinate $r$ by
\begin{eqnarray}
    r=r_s+ \eta\,  e^{i\Theta},
\end{eqnarray}
where $\eta$ is a  positive  real variable. In our case, it is convenient to choose  $\Theta=-\arctan(\omega_I/\omega_r)$ so that the expression (\ref{solinf}) becomes, using $r_*\simeq r$ when $r\to \infty$,
\begin{eqnarray}
 \lambda(\omega,r)\sim \lambda_{\rm in}e^{i (\omega r_s +\eta (\omega_R\cos(\Theta)-\omega_I\sin(\Theta)))}+ \lambda_{\rm out}e^{-i (\omega r_s +\eta (\omega_R\cos(\Theta)-\omega_I\sin(\Theta)))},
\end{eqnarray}
where we see that the exponential growth disappears from the solution. This result is used as a boundary condition at spatial infinity to integrate Eq.(\ref{pereq22}) from infinity to $\eta =0$, i.e. $r=r_s$. On the other hand, we integrate Eq.(\ref{pereq22}) from $r=0$ to $r=r_s$, imposing the regularity condition ($\lambda_-=0$). The quasinormal modess are obtained by matching the interior and  exterior solutions through the following equation:
\begin{eqnarray}
 \frac{d\lambda}{dr} \left(r=r_s\right)= e^{-i\Theta} \frac{d\lambda}{d\eta} \left(\eta=0\right),
\end{eqnarray}
which is satisfied only for discrete values  of $\omega$.

\begin{figure}[t]
\centering
\includegraphics[scale=1]{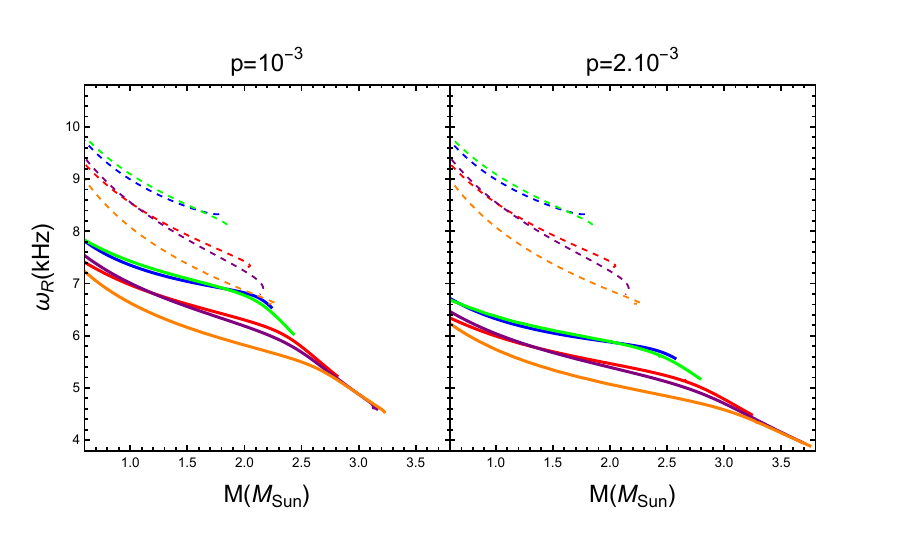}  
\caption{\small \small  
Real part $\omega_R$ of the fundamental $l=2$ QNM, as a function of the NS mass (in solar mass units), for the beyond-GR parameters $p=10^{-3}$ (left) and  $p=2\times 10^{-3}$ (right).
The GR ($p=0$) results are also plotted (dashed lines)  for comparison. Five distinct equations of state are considered: FPS (Blue), SLy (Orange), BSk19 (Green), BSk20 (Red) and BSk21 (Purple). }
\label{omega}
\end{figure}
\begin{figure}[t]
\centering
\includegraphics[scale=1]{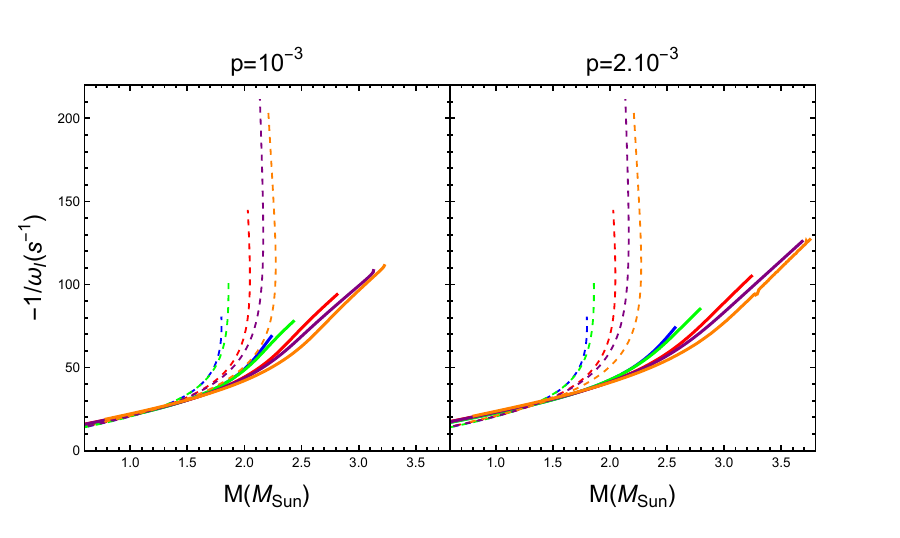}  
\caption{\small \small  
Damping time of the fundamental $l=2$ QNM, as a function of the NS mass (in solar mass units), for the beyond-GR parameters $p=10^{-3}$ (left) and  $p=2\times 10^{-3}$ (right), together with  the GR ($p=0$) results (dashed lines). Five distinct equations of state are considered: FPS (Blue), SLy (Orange), BSk19 (Green), BSk20 (Red) and BSk21 (Purple). 
}
\label{omega2}
\end{figure}

In Fig.~\ref{omega} and Fig.~\ref{omega2}, we have plotted the real part $\omega_R$  and the damping time $| \omega_I|^{-1}$ of the fundamental $l=2$  QNM, depending on the mass of the neutron star.  We see that, both in GR and in our model of modified gravity,  $\omega_R$ decreases as the neutron star mass increases, indicating that the oscillation frequency of the mode decreases for more massive stars. Meanwhile, Fig.~\ref{omega2} reveals that the imaginary part of  $\omega$ increases with mass, meaning that the damping time scale decreases, leading to shorter-lived oscillations for massive stars. This implies that as the neutron star becomes more massive, its oscillations not only slow down but also decay more rapidly. We also observe that  the star's internal structure, i.e. its equation of state, affects the  oscillation frequency and the damping time. Among the five equations of state we have studied,   the SLy equation of state has the lowest oscillations and highest damping time. 

The comparison between the left and right graphs in each figure allows us to see quantitatively how the modes are modified when the beyond-GR parameter $p$ increases. Note that the neutron masses can be much higher in our model than in GR, as discussed in \cite{Boumaza:2022abj}, which explains why the curves go much farther on the right in the modified gravity cases. We see that a higher value of $p$ leads to a significant decrease of the frequency $\omega_R$, whereas the influence of $p$ on the damping time is more moderate.
 
 Since both $p$ and the equation of state have an impact on the oscillations of the star, it might be difficult  to discriminate between the two. 
In order to lift this degeneracy, it is useful to try to identify universal relations for the (rescaled) frequencies and the (rescaled) damping times, i.e. relations that depend only  weakly  on the equation of state \cite{Tsui:2004qd,blazquez2022universal,Danchev:2020wnl}. Assuming a given
dependence of these quantities on the star compactness 
\begin{equation}
    C\equiv\frac{M}{r_s}\,,
\end{equation}
one can fit the numerical results obtained for a large number of star configurations (in practice, we use 115 different values of $C$ for each equation of state), with the central density $\rho_c$ varying in the interval $[4.6\rho_0,16.1\rho_0]$, where $\rho_0=m_n n_0=1.675\times 10^{-14}{ g}\cdot{\rm cm}^{-3}$, $m_n$ being the neutron mass and $n_0 = 0.1 {\rm  fm}^{-3}$  the typical number density in neutron stars. These  relations can be used to estimate the parameter $p$.   
If we assume a  quadratic dependence of $\omega_R r_s$ on the compactness $C=M/r_s$, we  obtain the following  phenomenological relations, for all the EOSs considered in this paper:
\begin{eqnarray}
\omega_R r_s=\left\{
        \begin{array}{ll}
            105.1(\pm 0.5)+6\,C(\pm 10 )-369 (\pm 22) C^2,&\qquad p=0 \ {\rm (GR)}\\
            98.6(\pm 0.9)+7\, (\pm 8) C - 273 (\pm 18) C^2 &\qquad p=10^{-3}\\
            97.5(\pm 0.7)-34\,(\pm 6)C  -  182.6 (\pm 14.2)C^2,&\qquad p=2\times 10^{-3}.
        \end{array}
    \right.
\end{eqnarray}
where the fitted coefficients change depending on the value  of  the parameter $p$.

\begin{figure}[h]
\centering  
\includegraphics[scale=0.89]{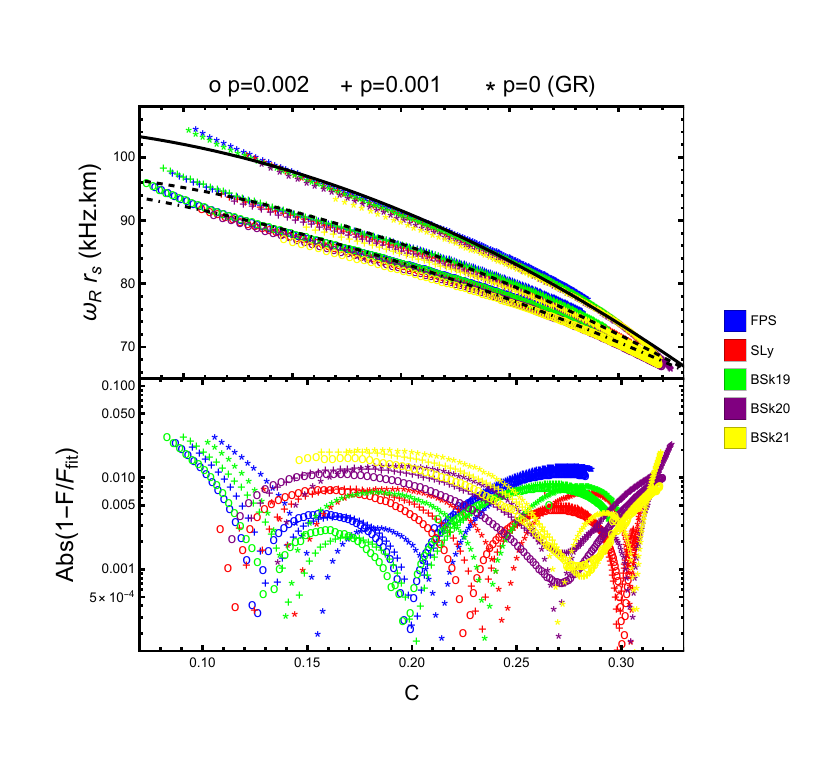}
\caption{\small \small  $\omega_R r_s$ as a function of the compactness for $p=0$ (continued line), $p=10^{-3}$ (dashed line), $p=2\times 10^{-3}$ (dot-dashed line).}
\label{wrc}
\end{figure}

We observe that the  term linear in $C$ is close to zero  for  GR and for the theory $p=10^{-3}$, which means it could be fixed to zero in practice for these cases. The coefficient of the linear term becomes   negative for the case $p=2\times 10^{-3}$. The deviation from the universal relation is less than $5\%$ in all cases,  similar to the  value obtained in other modified gravity models
\cite{blazquez2018axial,blazquez2016axial,blazquez2022universal,altaha2018axial}. As shown in Figure \ref{wrc}, when the parameter $p$ increases, the normalized frequency $\omega_R r_s$ decreases. The highest value corresponds to the GR case, while the lowest is for $p=2\times 10^{-3}$. 

A similar universal behaviour is found for the relation between $\omega_R M$ and   $C$, with different coefficients for the different values of $p$:
\begin{eqnarray}
\omega_R M=\left\{
        \begin{array}{ll}
           - 2.9(\pm 0.2) +111.5 (\pm 2.1) C -173.2 (\pm 4.7) C^2,&\qquad p=0 \ {\rm (GR)}\\
           - 1.9(\pm 0.2) +94.4 (\pm 1.6) C- 131.3 (\pm 3.5) C^2, &\qquad p=10^{-2}\\
           - 1.2(\pm 0.1) +84.4 (\pm 1.2) C - 108.0(\pm 2.6)  C^2,&\qquad p=2\times 10^{-2}\,.
        \end{array}
    \right.
\end{eqnarray}
\begin{figure}[h]
\centering
\includegraphics[scale=0.8]{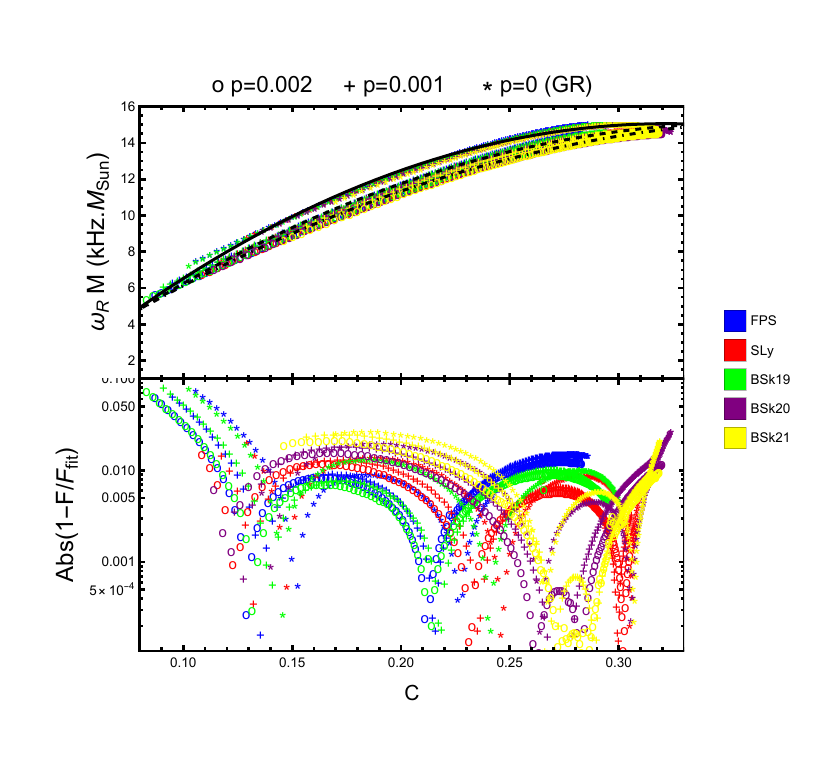}  
\caption{\small \small  Compactness-$\omega_R M$  relation for $p=0$ (continued lines), $p=10^{-3}$ (dashed lines), $p=2\times 10^{-3}$ (dot-dashed lines)  for various equations of state, distinguished by the colours.}
\label{wrm2}
\end{figure}
This universal relation looks well suited since the deviation is less than $5\%$ when the compactness of the star is in  the interval $[0.12,0.33]$, as shown  in Fig.~\ref{wrm2}, for all values of $p$. In addition, like in the previous relation $\omega_R M$ decreases by increasing the parameter $p$. The corresponding phenomenological $\omega_I M$-$C$ relation is given by
\begin{eqnarray}
\omega_I M=\left\{
        \begin{array}{ll}
            22.5(\pm 1.1) +356.4(\pm 9.7) C -1203 (\pm 21) C^2,&\qquad p=0 \ {\rm (GR)}\\
            32.9(\pm 1.1) +216.5(\pm 10.7) C -733 (\pm 23) C^2 &\qquad p=10^{-3}\\
            27.0(\pm 0.6) +262.7(\pm 6.1) C -803(\pm 14)  C^2,&\qquad p=2\times 10^{-3}.
        \end{array}
    \right.
\end{eqnarray}

\begin{figure}[h]
\centering
\includegraphics[scale=0.8]{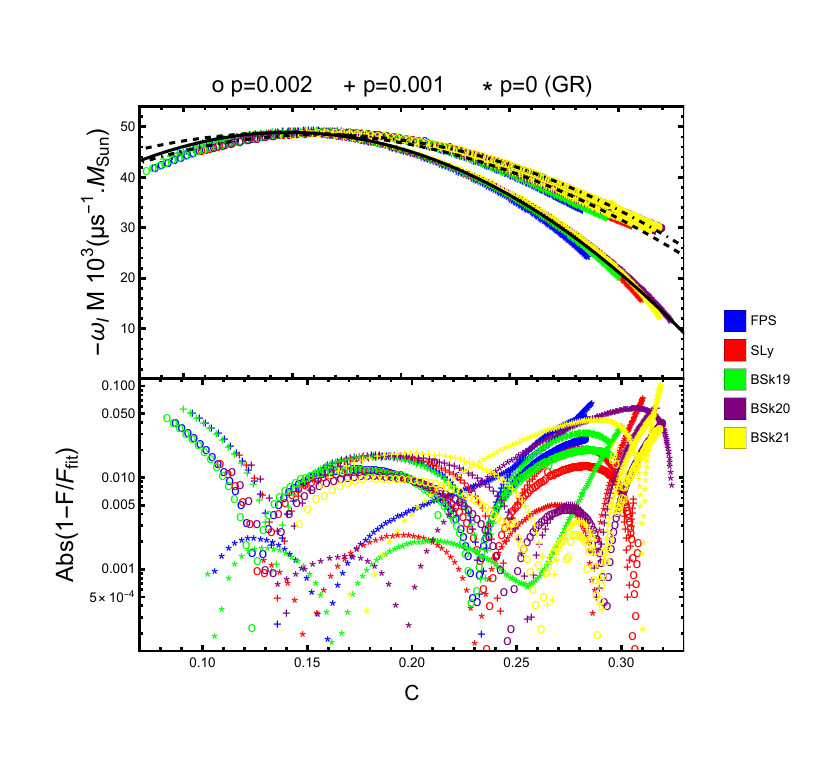}  
\caption{\small \small  $M \omega_I$ as a function of the compactness for $p=0$ (continued lines), $p=10^{-3}$ (dashed lines), $p=2\times 10^{-3}$ (dot-dashed lines).}
\label{pwi}
\end{figure}
An alternative phenomenological universal   relation \cite{Blazquez-Salcedo:2018tyn,blazquez2022universal} is obtained by rescaling the real part and imaginary part of the quasinormal modes  as  $\tilde{\omega}_I=\omega_I/\sqrt{P_c}$ and  $\tilde{\omega}_R=\omega_R/\sqrt{P_c}$, where $P_c$ is the central pressure. The relation between the two is described by expressions of the form
\begin{eqnarray}
\tilde{\omega}_I =\left\{
        \begin{array}{ll}
            -0.379(\pm 0.016) +0.400(\pm 0.011) \tilde{\omega}_R +0.082 (\pm 0.002) \tilde{\omega}_R^2,&\qquad p=0 \ {\rm (GR)}\\
            -0.151(\pm 0.017) +0.374(\pm 0.015) \tilde{\omega}_R +0.124 (\pm 0.003)  \tilde{\omega}_R^2, &\qquad p=10^{-3}\\
            -0.174(\pm 0.015) +0.456(\pm 0.015) \tilde{\omega}_R +0.144 (\pm 0.003)   \tilde{\omega}_R^2,&\qquad p=2\times 10^{-3}.
        \end{array}
    \right.
\end{eqnarray}
\begin{figure}[h]
\centering
\includegraphics[scale=0.8]{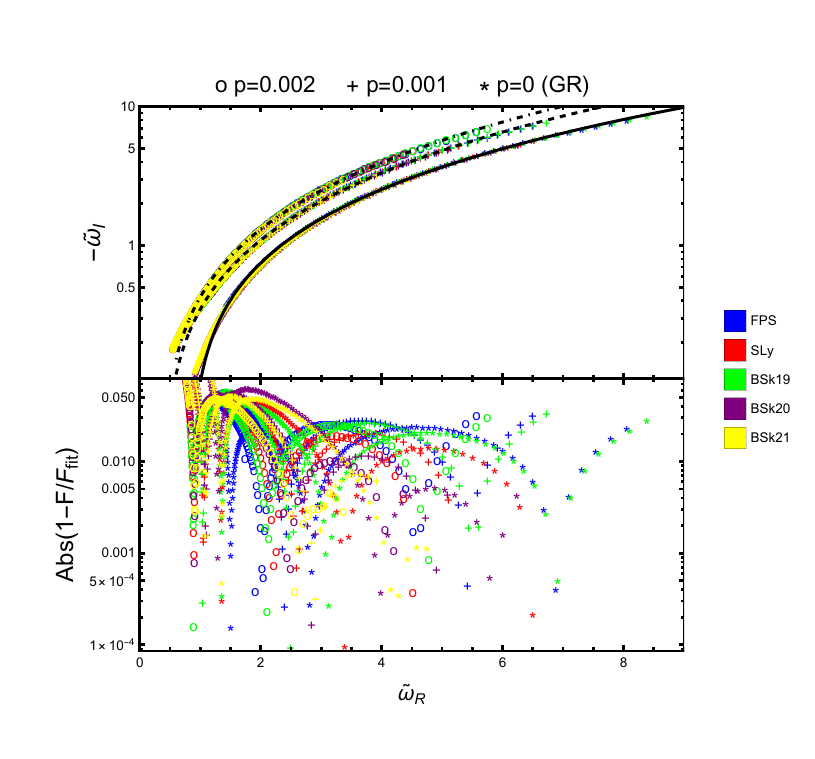}  
\caption{\small \small $\tilde{\omega}_R $-$\tilde{\omega}_I$ relation for $p=0$ (continued line), $p=10^{-3}$ (dashed line), $p=2\times 10^{-3}$ (dot-dashed line).}
\label{pwiwr}
\end{figure}
These relations provide a good fit for our model, with deviations remaining within $10\%$ in all cases, as illustrated in Fig~\ref{pwiwr}. We observe that the quadratic coefficient increases significantly when the parameter $p$ increases  which indicates that $\tilde{\omega}_I$ is more sensitive to the deviation from GR  for higher  values of $\tilde{\omega}_R$.

\section{Conclusion}\label{conclusion}

In this paper, motivated by the exploration of possible deviations from general relativity  in the strong-field regime, we have studied the axial quasinormal modes  of neutron stars in a specific subclass of DHOST theories. As a starting point,  we have used  the   static and spherically symmetric stellar configurations computed in our previous paper~\cite{Boumaza:2022abj}   by solving the   modified TOV equations with realistic equations of state. We have then obtained the dynamics of the axial perturbations by expanding  the total action to second order in the axial sector. This enabled us to show the absence of  ghost or gradient instabilities for the (single) degree of freedom associated with the axial mode. We then derived the equation of motion governing axial perturbations, which leads, with appropriate boundary conditions, to the QNMs  of the neutron star.

Interestingly, we have noticed that, for axial modes, the correspondence between the DHOST dynamics and the usual GR one in an effective metric, discussed recently in the context of DHOST black holes \cite{Langlois:2022ulw} (see also \cite{Noui:2023ksf,Charmousis:2025xug}), is still valid for neutron stars. This can be understood by the fact that  axial modes are decoupled from the fluid modes  and therefore are only sensitive to the background geometry.  

In our model, we have found that the fundamental $l=2$ QNM exhibits a deviation from its GR counterpart in both frequency and damping time. But the QNMs are also sensitive to the choice of the equation of state for the neutron star matter. In order to disentangle the impact of modified gravity from that of the equation of state,   we have looked for universal relations and identified  empirical relation between the real and  imaginary parts of the fundamental QNM  and the compactness of the star. Our results show that these relations can allow us to discriminate  between GR and our DHOST model independently of the kind of equation of state. 

Overall, our findings highlight the relevance of neutron star spectroscopy   to probe alternative theories of gravity. In future work, it would be interesting to extend the present analysis to polar perturbations, which are  coupled to the fluid modes.

\newpage
\appendix

\section{Variational principle for a perfect fluid}
\label{Appendix_fluid}
\subsection{Action and equations of motion}
In this appendix, we  present one of the variational principles for perfect fluids in general relativity, due to Schutz~\cite{schutz1970perfect}.
It is based on the following action:
\begin{eqnarray}
S_m[L,A,B] =\int d^4 x\sqrt{-g}\, P(\mu) \,, 
\end{eqnarray}
where $\mu$ depends on the three scalar fields $L$, $A$ and $B$, trough the definition 
 \begin{eqnarray}
\mu= \sqrt{-g^{\alpha\beta}\left(\partial_\alpha L+ A\partial_\alpha B\right)\left(\partial_\beta L + A\partial_\beta B\right)}\,.\label{tmu22}
 \end{eqnarray}
Physically, $\mu$  corresponds to the chemical potential,  $P(\mu)$ is the pressure, and the 4-velocity of the fluid is given by the time-like unit vector 
\begin{eqnarray}\label{uschutz}
u_\alpha = \frac{1}{\mu}\left(\partial_\alpha L + A\partial_\alpha B\right)\,.
\end{eqnarray}
In the special case of an   irrotational fluid, it  is sufficient to use a single scalar field and the scalar fields $A$ and $B$ can be ignored. 

Let us now check that the action \eqref{action} indeed leads to the usual equations of motion for a relativistic perfect fluid. 
  Varying (\ref{action}) with respect to $L$, $A$ and $B$, one finds, respectively, 
\begin{eqnarray}\label{cnstrntaB}
\nabla_\alpha\left(\frac{dP}{d\mu} u^\alpha\right)=0,\qquad u^\alpha\partial_\alpha A=0,\qquad u^\alpha\partial_\alpha B=0.
\end{eqnarray}
One recognises in the first equation of \eqref{cnstrntaB} the usual conservation law $\nabla_\alpha(n u^\mu)=0$, where $n=dP/d\mu$ is the number density. Then, using the relation 
\begin{equation}
\nabla_{[\alpha}(\mu u_{\beta]})=\partial_{[\alpha}A\ \partial_{\beta]}B\,,
\end{equation}
which follows from \eqref{uschutz}, and taking into account the second and the third equations in \eqref{cnstrntaB}, we obtain the equation
\begin{eqnarray}\label{deltamu}
u^\alpha\nabla_{[\alpha}(\mu u_{\beta]})=0\,,
\end{eqnarray}
which corresponds to the relativistic Euler equation for a perfect fluid.

Finally, the variation of the action \eqref{action} with respect to the metric $g_{\alpha\beta}$ yields the usual energy-momentum
\begin{eqnarray}
T_{\alpha\beta}=(\rho+P)u_\alpha u_\beta + P g_{\alpha\beta}\,,
\end{eqnarray}
after using the expression \eqref{rho} for the energy density $\rho$.

\subsection{Background fluid}
In the background, the fluid is at rest so its spatial components vanish. The time component is determined by the normalisation condition $g_{\alpha\beta}u^\alpha u^\beta =-1$. For the metric  (\ref{ds}), this gives 
\begin{eqnarray}
\label{u_bckgd}
\Bar{u} ^\alpha = \left\lbrace \frac{1}{\sqrt{f}},0,0,0 \right\rbrace\,, \qquad \Bar{u}_\alpha = \left\lbrace -\sqrt{f},0,0,0 \right\rbrace\,. 
\end{eqnarray}
In this case, the fluid is irrotational and one can ignore the scalars $A$ and $B$.
Therefore, integrating the component $t$ of Eq.(\ref{uschutz}) with respect to $t$, gives
\begin{eqnarray}
\label{L_bckgd}
L(t,r)&=& -\sqrt{f}\mu t.
\end{eqnarray}
Substituting this result in the component $r$ of  Eq.(\ref{uschutz}), we obtain, since $u_r=0$, the following constraint:
\begin{eqnarray}\label{muequation}
\left(\mu\sqrt{f}\right)'=0\quad {\rm or}\quad\frac{\mu'}{\mu}&=& -\frac{f'}{2 f}\,.
\end{eqnarray}
 Finally, the matter conservation equation (\ref{e4}) can be recovered by multiplying (\ref{muequation}) by $(dP/d{\mu})$ and using the formulas $P'=\mu'(dP/d{\mu})$ and \eqref{rho}.

\subsection{Quadratic matter action for the axial perturbations}

The action at quadratic order in the metric and fluid perturbations is useful to
 determine the dynamics of the  linear perturbations. In the following, we explain how  the matter action  at
 quadratic order is computed.

 By expanding, up to second order, the relation \eqref{uschutz}, using the background values \eqref{u_bckgd} and \eqref{L_bckgd} and the perturbations \eqref{perturbed_metric} and \eqref{4-velocity}, 
  we find\footnote{We expand the scalar field $L$ up to second order, but the scalar fields $A$ and $B$ only up to first order since they vanish in the background and appear quadratically in \eqref{uschutz}.} that the first order perturbation of $\mu$ vanishes while the second order perturbation is given by
\begin{eqnarray}
\delta^2 \mu(t,r,\theta)&=& -\frac{r^2\mu}{2 f }   v^2 \,(\partial_\theta Y_{l0})^2\,.
\end{eqnarray}
Moreover, the expansion of  $\sqrt{-g}$ yields, at second order, the expression
\begin{eqnarray}
 \delta^2 \sqrt{-g}=\frac12 r^4 \left(\sqrt{\frac{h}{f}}H_0^2-\sqrt{\frac{f}{h}} H_1^2\right)\sin{\theta}\,(\partial_\theta Y_{l0})^2\,.
\end{eqnarray}
Substituting the above expressions into  the second order matter action, we get 
\begin{eqnarray}
S_{\rm m,axial} &=&\int d^4x\left( P \delta^2 \sqrt{-g} +\sqrt{-g}\frac{dP}{d\mu} \delta^2 \mu\right)
\\
&=&-\frac{l(l+1)}{2}\int drdt \,\sqrt{f h}\,r^4\left(\frac{ P }{ h}H_1^2+\frac{\rho }{ f}H_0^2\right)\,,
\end{eqnarray}
where we have replaced $v$ by $-H_0$, according to the constraint \eqref{H0}, used the identity \eqref{rho} and integrated over the angles $\theta$ and  $\phi$.

\section{Odd-Parity perturbations}
\label{Appendix_axial}
We consider a linearly perturbed metric 
\begin{equation}
    g_{\mu\nu}=g^{0}_{\mu\nu}+h_{\mu\nu}
\end{equation}
$g^{0}_{\mu\nu}$ is a static and spherically symmetric metric, of the form \eqref{ds}, and $h_{\mu\nu}$ is a  small perturbation,  i.e.  $ |h_{\mu\nu}|\ll|g^{(0)}_{\mu\nu}|$. Since the background is spherically symmetric, one can decompose the perturbations  into odd and even parity components, which are fully decoupled at the linear level. 

Focusing now on the odd-parity (or axial) sector, it is convenient to decompose  the perturbations $h_{\mu\nu}$  into  spherical harmonics $Y_{lm}(\theta,\phi)$, which gives in general
\begin{eqnarray}
    h_{at}&=& r^2\sum_{lm} H_{0}^{lm}(r,t)\sin\theta\,\epsilon_{ab}\nabla^b Y_{lm}\\
    h_{ar}&=&r^2\sum_{lm} H_{1}^{lm}(r,t)\sin\theta\,\epsilon_{ab}\nabla^b Y_{lm},\\
 h_{ab}&=&\frac{1}{2}r^2 \sum_{lm} Q^{lm}(r,t)\sin\theta(\epsilon_{a}^{c}\nabla_c\nabla_b Y_{lm}+\epsilon_{b}^{c}\nabla_c\nabla_a Y_{lm}),
\end{eqnarray}
where $\{a,b\}\equiv\{\theta,\phi\}$;  the indices are raised or lowered using the unit 2-sphere metric and $\epsilon_{ab}$ is the antisymmetric tensor with $\epsilon_{\theta\phi}=-1$.
Note that the presence of $\epsilon_{ab}$ in the above expressions characterises the axial modes, in contrast with the polar (or even-parity) modes.

By performing an infinitesimal gauge transformation $x^\mu\rightarrow x^\mu+\xi^\mu$, specialised to the axial sector, i.e. of the form  $\xi^t=\xi^r=0$ and 
\begin{eqnarray}
    \xi^a=r^2\sum_{lm} \Lambda^{lm}(r,t)\sin\theta\,\epsilon^a_{\  b}\nabla^b Y_{lm},
\end{eqnarray}
 the functions $H_{0}(r,t)$, $H_{1}(r,t)$ and $Q(r,t)$ are modified as follows:
\begin{eqnarray}
    H_{0} & \rightarrow  H_{0}+\dot{\Lambda},\quad
    H_{1}  \rightarrow  H_{1}+\Lambda',\quad
    Q  \rightarrow  Q+2\Lambda\,,
\end{eqnarray}
where we have omitted  the indices $l$ and $m$ to lighten the notation. 
 As is standard, we have fixed  the gauge such that $Q^{lm}=0$, which is always possible by choosing the appropriate $\Lambda^{lm}$ that cancels an initial non zero value for $Q^{lm}=0$. In this gauge, the axially perturbed  metric reduces to \eqref{perturbed_metric} for $m=0$, the  other modes being degenerate with the mode $m=0$ as a consequence of the spherical symmetry of the background.

Finally, let  us mention that the axial perturbation of the four-vector velocity, $\delta u^\mu$, has the same form as the gauge transformation vector $\xi^\mu$:  the nonzero components can be decomposed as 
\begin{eqnarray}
  \delta u^{a}=r^2 f^{-1/2}\sum_{lm} v^{lm}(r,t)\sin\theta\,\epsilon^{a}_{\  b}\nabla^b Y_{lm},
\end{eqnarray}
where we have introduced the normalisation factor $f^{-1/2}$ for convenience.

\bibliographystyle{utphys}
\bibliography{DHOST_NS_biblio1}

\end{document}